# Crystallogprahically selective nanopatterning of graphene on SiO$_2$

*Péter Nemes – Incze[*,1], Gábor Magda[2], Katalin Kamarás[3] and László Péter Biró[1]*

*1* Research Institute for Technical Physics and Materials Science, H-1525 Budapest, P.O. Box 49, Hungary,

*2* Budapest University of Technology and Economics, H-1521 Budapest, PO Box 91, Hungary,

*3* Research Institute for Solid State Physics and Optics, Hungarian Academy of Sciences, H-1525 Budapest, P.O. Box 49, Hungary

**Graphene has many advantageous properties [1], but its lack of an electronic band gap makes this two dimensional material impractical for many nanoelectronic applications, for example field effect transistors [2]. This problem can be circumvented by opening up a confinement induced gap, through the patterning of graphene into ribbons having widths of a few nanometres. The electronic properties of such ribbons depend on their size and the crystallographic orientation of the ribbon edges [3, 4]. Therefore, etching processes that are able to differentiate between the zigzag and armchair type edge terminations of graphene are highly sought after. In this contribution we show that such an anisotropic, dry etching reaction is possible and we use it to obtain graphene ribbons with zigzag edges. We demonstrate that the starting positions for the carbon removal reaction can be tailored at will with precision.**

---

[*] Corresponding author; e-mail: nemes@mfa.kfki.hu, Phone: +36-1-3922526, Fax: +36-1-3922226, www.nanotechnology.hu

Graphene structures just a few nanometres in size, such as ribbons or triangles are predicted to have different electronic properties depending on the orientation of the edges [3-5]. For example, graphene ribbons with zigzag type edge termination are presumed to exhibit spin polarized states at the ribbon edges [6, 7], which give rise to magnetic properties like large magnetoresistance [8] and the possibility to be applied in spintronic devices. However, the experimental investigation of these kinds of devices is not yet possible, because there is a lack of a method to produce such crystallographically oriented structures in a controlled fashion. The STM lithography of graphene developed in our group gives us the ability to cut graphene nanoribbons of arbitrary orientation, but only on a conductive surface [9]. Several theoretical investigations show that the armchair and zigzag edges of graphene have differing reactivity [10, 11] and energetic stability [12], pointing to the possibility of anisotropic etching. Indeed, recently the crystallographic orientation dependent etching of graphite by metal nanoparticles [13] has been successfully implemented on graphene samples [14]. In the case of graphite etching, Ci et al. have shown evidence that zigzag edges are produced [15]. However, up to now control over the patterning has not been demonstrated, as the catalytic particle trajectories cannot be guided at will. Below we present a controllable and anisotropic etching process which circumvents all these problems.

For all our patterning experiments the graphene samples were prepared by micromechanical cleavage [16], supported on single crystal silicon wafers having a 90 nm thick [17] $SiO_2$ top layer. After preparing graphene samples this way, we have exposed them to an oxygen – nitrogen atmosphere at $500^\circ C$ (see online Supporting Material). This treatment produces circular etch pits on the graphene surface [18], as can be seen in Fig. 1a. This was followed by a subsequent etching step, which consisted of annealing the sample under a continuous flow of Ar gas at $700^\circ C$. After this second treatment step, the existing circular etch pits continued to grow in size but we observed that they have now transformed into hexagonal pits and importantly: no new etch pits have been formed (see Fig 1b). The fact that the etch pits are hexagonal, as opposed to circular and that they all have the same orientation relative to one another means that carbon removal from either zigzag of armchair edges has a very different reaction rate [19]. Examining the AFM images of the etch pits more closely reveals that roughly 1 nm of the silica substrate is missing, where the graphene has receded during the 700$^o$C annealing step. This shows up as a depression of the substrate beside the edges of the hexagonal holes (see Fig. 1b and the line cut in Fig. 2a). In light of this, we propose that the main mechanism for the formation of the hexagonal pits is the carbothermal reduction of the substrate $SiO_2$ by the carbon in the graphene edges:

$$SiO_2 + C \stackrel{700^oC}{=} SiO \uparrow + CO \uparrow$$

The above reaction needs temperatures higher than 1754$^o$C to proceed under equilibrium conditions, at atmospheric pressure [20], but it has been demonstrated to occur at much lower temperatures. For example Byon and Choi [21] have taken advantage of the same carbothermal reaction at 830$^o$C and have shown that carbon nanotubes can be used as a guide and carbon source for the etching of nanotrenches into $SiO_2$ surfaces. Both reaction products of the carbothermal etching: CO and SiO, are volatile at 700 $^o$C and thus get swept away by the Ar gas flow, making possible the very slow removal of carbon. The fact that the edges are so well formed (see Fig. 2a) shows that the above reaction has a high crystallographic selectivity under these conditions i.e., it has a much higher reaction rate for the removal of carbon from one type of graphene edge than from the other. The edge roughness of the hexagonal holes cannot be precisely determined from AFM images, due to tip convolution effects, but it is clearly of the order of nanometres. Theoretical studies of armchair and zigzag edges have shown that they do have different reactivities [10, 11], and different chemical processes have affinity to etch different edge orientations [15, 19].

To reveal the orientation of the hexagonal hole edges, we have prepared metallic contacts [22] to our samples and have performed atomic resolution STM measurements. After matching the STM images to the AFM images of the etch pits, we were able to show that the edge orientation of these pits is of zigzag type (see Fig. 2c). Further details on the STM measurement can be found in the online Supporting Material. We have to emphasize at this point that the STM measurements were obtained not exactly at the hexagonal pit edges, but a few hundred nanometres further away from them (see Supp. Mater.), making possible the identification of the crystallographic axes of the sample and therefore the edge orientation, but without any information on the structure of the edge itself. Measuring exactly at the hexagonal hole edge with STM is rather difficult, because of the transition from the graphene to the insulating $SiO_2$. Further investigation is needed to elucidate the magnitude of the edge roughness and exact structure of the edges through STM and STS investigation, as edge disorder [23], reconstruction of the edge [12] and/or functionalisation [7, 24] will very likely dramatically influence the properties of the devices. Indeed, by following the evolution of the Raman spectra acquired on the samples annealed at 700$^o$C, with prior oxidation (Fig 2b) and without prior oxidation (Supporting Mater.) we can observe a significant upshift of the G and 2D peaks, normally found at 1580 cm$^{-1}$ and 2700 cm$^{-1}$ respectively. Such changes in the Raman frequencies are most likely due to a convolution of two effects. According to the work of Das et al. this upshift of about 20 cm$^{-1}$ can be attributed to a strong hole doping of the graphene layer [25] and has also been observed by Liu et al [18]. From the electrochemical top-gate doping experiments of Das et al. we can estimate the Fermi level shift to be aroud 0.48 eV with a hole density of $1.5 \times 10^{13}$ cm$^{-2}$. Graphene samples prepared under ambient conditions usually show p doping due to adsorbates [26], but in our case such high doping levels most likely originate from the attachment of oxygen species at the highly reactive graphene edges during etching [7]. Our Raman data correlate nicely with observations of Das et al. (see supporting information). However, another contribution to the observed G and 2D shifts in the Raman spectra could be due to compressive stress of the graphene layers after thermal cycling due to the very different thermal expansion coefficients of graphene and its substrate [31, 32].

Another important feature of the Raman spectra is the absence of the D peak of graphene around 1350 cm$^{-1}$ from both the pristine and etched samples. This suggests that although the graphene layer gets etched out at the edges, it retains its nearly perfect crystal structure. Further findings supporting this conclusion are the atomic resolution STM images obtained at multiple locations on the sample annealed at 700$^{o}$C for 30 minutes. These showed only the unperturbed atomic resolution image of graphene (Fig. 2c), with no sign of $\sqrt{3}\times\sqrt{3}R30^0$ type reconstruction, the typical hallmark of defect scattering [27]. These results support the conclusion that during carbothermal etching, the dissociation of carbon bonds only occurs at the sample edges where the carbon atoms have less than three neighbours and the binding energy of the atoms is lower than for lattice positions having three nearest neighbours. This has profound consequences regarding the controllability of the etching process.

In the experiments described above the hexagonal etch pits were grown in the places defined by the circular oxidation pits. This is a random network of native defects, which gives little control over the obtained hexagonal hole distribution and architecture. A major advantage of the carbothermal etching process is that the distribution of the hexagonal holes is directly linked to the sample defect distribution, so by tailoring the defect positions one should be able to control the arrangement of the pattern composed of hexagonal holes. The pre-patterning of defect sites may be achieved in many ways, through e-beam- or photo-lithography for example. We chose another simple patterning technique, using the tip of an AFM probe as an indentation tool. During AFM indentation one can puncture the graphene lattice at predefined positions and thus create and ordered array of defects (see Fig. 3a). The benefit of using AFM is its precision in the X-Y direction and the ability to quickly pattern periodic structures onto the graphene without the issue of resist contamination. An example of this can be seen in Fig. 3a, where we have prepared a 3×3 matrix of indentation holes in graphene. Then by annealing the sample, hexagonal pits can be grown starting from the puncture holes introduced by the AFM tip (Fig. 3b). This structure corresponds to a graphene antidot lattice [28], but with crystallographically oriented hexagonal lattice points. The size of these hexagonal pits can be adjusted by changing the annealing time. This is nicely illustrated in Fig. 4, where we present three etch pits prepared using AFM indentation, after successive 2 hour carbothermal etch cycles. Starting out with etch pits of 300 nm size, the pits can be grown in size by 50 nm after each 2 hour etching cycle. This gives a growth rate of about 12 nm/hour, allowing for a good control of pit dimensions.

Having this high controllability allows us to use the hexagonal holes as nanosized building blocks to assemble graphene devices such as nanoribbons. An example of a nanoribbon can be seen in Fig. 5, where the device is produced between two etch pits. The nanoribbon has zigzag edges and a width of 35 nm; such ribbons of graphene already show confinement effects [3, 29]. To the best knowledge of the authors this is the first example of a graphene nanoribbon, with well defined zigzag orientation of the edges, being produced on an insulating substrate, using a well controlled patterning process. In addition, more complex architectures, such as nanoribbon "Y" junctions may also be obtained without much difficulty (see Fig. 5). A further important advantage of the controlled oxidation is that it very clearly reveals the grain boundaries (see Fig. 1), thus it gives us the possibility to avoid the production of devices placed over two neighbouring grains, where the grain boundary would certainly affect the transport properties.

The carbothermal etching process described here could be coupled with other lithography techniques, like electron beam lithography or arrays of AFM tips [30] to produce graphene nanostructures with zigzag orientation of the edges. The graphene retains its high crystallinity after the etching treatment and the process makes possible the fabrication of very narrow ribbons of graphene, down to 35 nm and possibly even below. We need to stress that the materials and temperatures used to achieve the etching are completely compatible with semiconductor industry processes, helping to achieve the integration of crystallographically oriented graphene nanostructures into existing silicon electronics.

**ACKNOWLEDGMENT** Financial support by OTKA-NKTH grants 67793, 67851 and 67842 is acknowledged.

[27] Tapasztó, L.; Dobrik, G.; Nemes - Incze, P.; Vertesy, G.; Lambin, P.; Biró, L. P. Tuning the electronic structure of graphene by ion irradiation. *Phys. Rev. B*. **2008**, *78*, 233407.

[28] Eroms, J.; Weiss, D. Weak localization and transport gap in graphene antidot lattices. *New Journal of Physics*. **2009**, *11*, 095021.

[29] Masubuchi, S.; Ono, M.; Yoshida, K.; Hirakawa, K.; Machida, T. Fabrication of graphene nanoribbon by local anodic oxidation lithography using atomic force microscope. *Appl. Phys. Lett.* **2009**, *94*, 082107.

[30] Knoll, A.; Bächtold, P.; Bonan, J.; Cherubini, G.; Despont, M.; Drechsler, U.; Dürig, U.; Gotsmann, B.; Häberle, W.; Hagleitner, C. Integrating nanotechnology into a working storage device. *Microelectron. Eng.* **2006**, *83*, 1692–1697.

[31] C. Chen, W. Bao, J. Theiss, C. Dames, C.N. Lau, S.B. Cronin, Raman Spectroscopy of Ripple Formation in Suspended Graphene. *Nano Lett.* **2009**. *in press*, DOI: 10.1021/nl9023935

[32] G. Tsoukleri, J. Parthenios, K. Papagelis, R. Jalil, A.C. Ferrari, A.K. Geim, et al., Subjecting a graphene monolayer to tension and compression. *Small* **2009**, *5*, 2397-402.

**Figures and Captions**

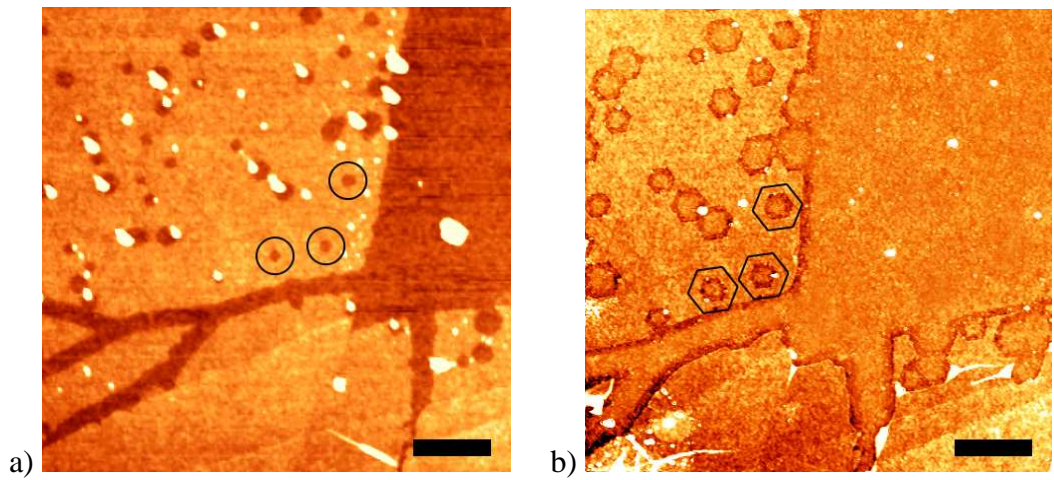

**Figure 1.** Image of a graphene layer oxidized in a mixture of O2 and N2 gas, at 500°C for 40 minutes (a) and the subsequent growth of hexagonal holes in the graphene, after annealing at 700°C in an Ar atmosphere (b). Specific oxidation holes are marked by black circles, the same etch pits are marked by hexagons after carbothermal etching. The long, branching etch marks are due to the etching out of grain boundaries. Scale bars are 1 µm. White spots on the AFM images are the remnants of the scotch tape material and have no effect on the oxidation or annealing processes (for more information see ref. **Hiba! A könyvjelző nem létezik.**).

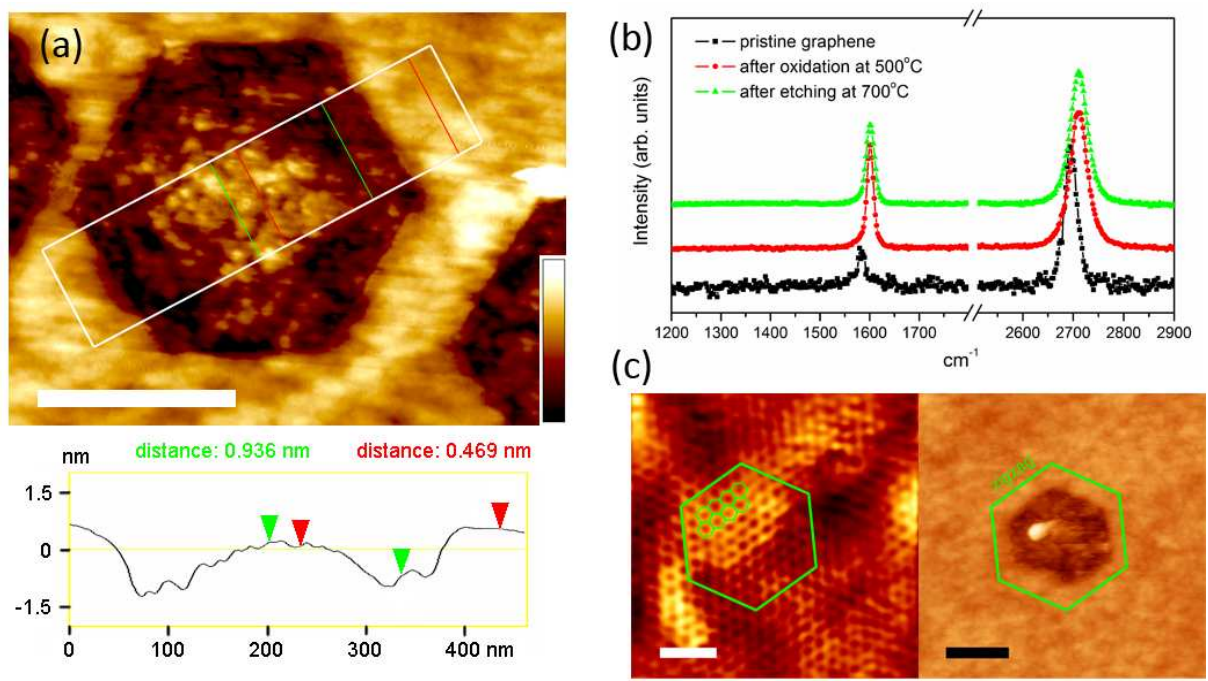

**Figure 2.** (a) AFM image of a hexagonal etch pit from the sample shown in Fig. 1. In the height profile we can clearly identify that the SiO$_2$ substrate has been etched away along with the graphene at the oxidation hole edges. (colour bar 3 nm, scale bar 200 nm). (b) Raman spectra of pristine, oxidized and annealed graphene. (c) STM image (scale bar 1 nm) and AFM image (scale bar 150 nm) side by side, showing the atomic lattice of the graphene sample and the orientation of the etch pit. The hexagons help to visualize the orientation relationship between the two images.

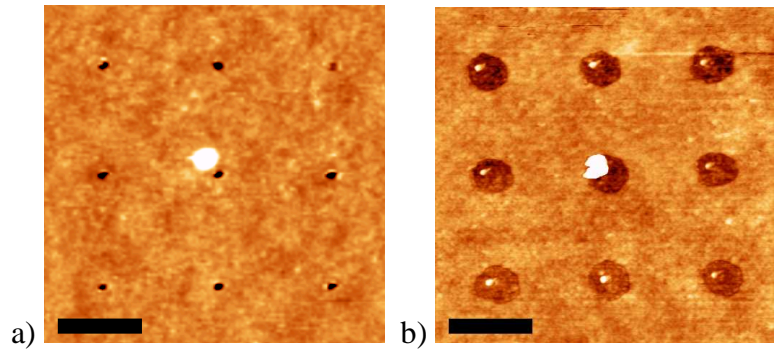

**Figure 3.** The result of AFM indentation: a 3×3 matrix of holes in graphene (a). AFM micrograph of the hexagonal holes grown from the defects induced by indentation (b). Scale bars 500 nm. Protrusions inside the hexagons are remnants of the AFM indentation process.

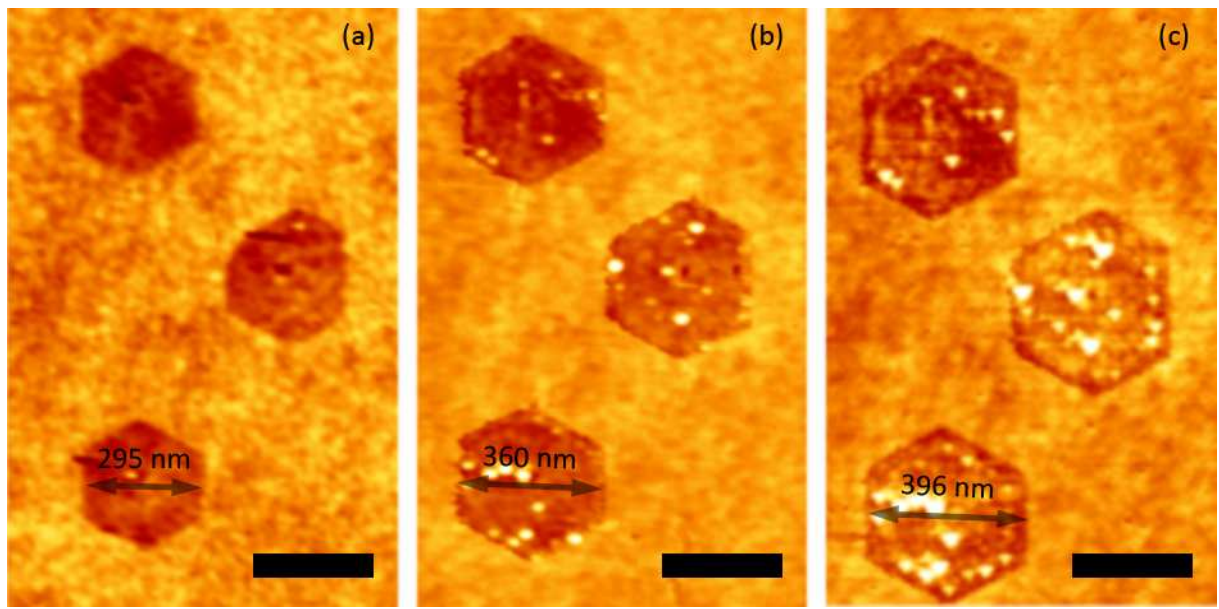

**Figure 4.** AFM images showing the controllability of carbothermal etching. Scale bars 300 nm. a) Three hexagons produced by AFM indentation of about 300 nm size. Images b) and c) show the same etch pits after an additional 2 and 4 hours of etching respectively.

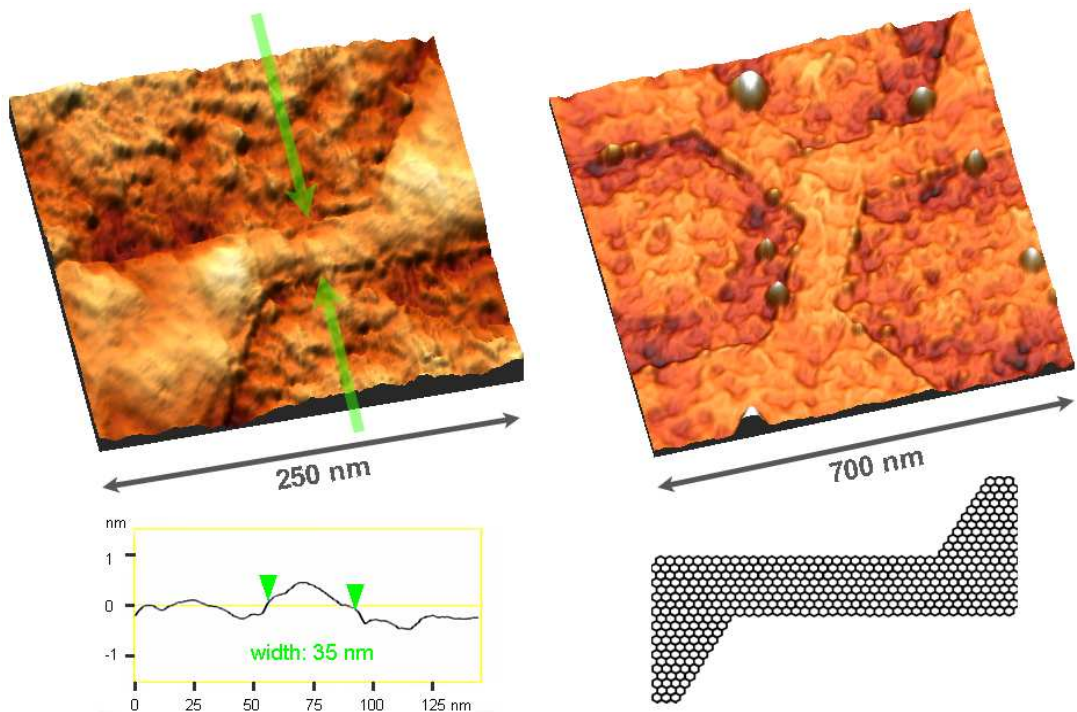

**Figure 5.** The left image shows a 3D AFM image of a graphene nanoribbon of about 35 nm width, which can be also seen in the top left corner of Fig. 2. The AFM height profile has been acquired at the place shown by green arrows. The inset in the right lower corner shows a scheme of the corresponding atomic structure. The right image shows a junction of 3 nanoribbons, with the ribbons having widths of: 93, 100, 101 nm (starting from the upper left ribbon, going clockwise).

**Table of Contents Graphic**

Graphene nanoribbon, supported on $SiO_2$, fabricated by carbothermal etching.

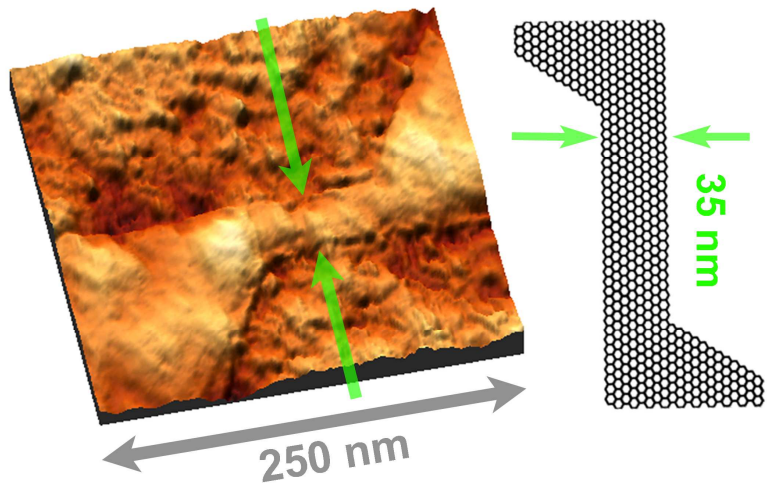